\documentclass[twocolumn,english,aps,prl,floatfix,tightenlines]{revtex4-1}
\usepackage[T1]{fontenc}
\usepackage[latin9]{inputenc}
\usepackage{amsmath}
\usepackage{amssymb}
\usepackage{graphicx}

\makeatletter

\@ifundefined{textcolor}{}
{%
 \definecolor{BLACK}{gray}{0}
 \definecolor{WHITE}{gray}{1}
 \definecolor{RED}{rgb}{1,0,0}
 \definecolor{GREEN}{rgb}{0,1,0}
 \definecolor{BLUE}{rgb}{0,0,1}
 \definecolor{CYAN}{cmyk}{1,0,0,0}
 \definecolor{MAGENTA}{cmyk}{0,1,0,0}
 \definecolor{YELLOW}{cmyk}{0,0,1,0}
}

\usepackage{babel}
\usepackage{times}
\usepackage{hyperref}
\usepackage{url}

\newcommand{\tr}{\mathrm{Tr}}

\newcommand{\1}{\leavevmode{\rm 1\ifmmode\mkern  -4.8mu\else\kern -.3em\fi I}}

\makeatother

\usepackage{babel}
\begin{document}

\title{Dynamical Response Theory for Driven-Dissipative Quantum Systems}

\author{Lorenzo Campos Venuti and Paolo Zanardi}

\affiliation{Department of Physics and Astronomy, and Center for Quantum Information
Science \& Technology, University of Southern California, Los Angeles,
CA 90089-0484}
\begin{abstract}
We discuss  dynamical response theory of driven-dissipative quantum systems
described by Markovian Master Equations generating semi-groups of maps.
In this setting thermal equilibrium states are replaced by non-equilibrium
steady states and  dissipative perturbations are considered besides the Hamiltonian ones. We derive  explicit
 expressions for  the linear dynamical response functions for  generalized dephasing 
channels and for Davies thermalizing generators. We introduce the notion of maximal harmonic response
and compute it exactly for a single qubit channel. 
Finally, we analyze linear response near dynamical phase
transitions in quasi-free open quantum systems. It is found that the
effect of the dynamical phase transition shows up in a peak at the
edge of the spectrum in the imaginary part of the dynamical response function. 
\end{abstract}
\maketitle

\section{Introduction}

Computing the response of an observable expectation value to a small
time-dependent perturbation is one of the most successful way to relate
physical quantities to the underlying theoretical description of the
system. In this way one can relate various fundamental quantities
such as electric or heat conductivity, magnetic susceptibilities,
Hall conductance and so on, to microscopic properties of the underlying
physical model. The classical paper of Kubo \citep{kubo_statistical-mechanical_1957}
gives formulae to compute such dynamical susceptibilities for a closed
quantum mechanical systems ``not far apart from thermal equilibrium''.
In Ref.~\citep{kubo_statistical-mechanical_1957} the system is supposed
to have reached, by some mean, a thermal equilibrium state which gets
slightly modified under the effect of the external perturbation. In
order to reach such equilibrium state presumably an interaction with
an external environment was crucial. The effect of the environment
is however considered small and, in fact, completely neglected. The
system is then supposed to evolve isolated from the environment according
to Schr\"{o}dinger equation. 

Kubo formulae have since been utilized countless times (for a beautiful
example consider the quantization of Hall conductance in topological
insulator, see e.g.~\citep{hasan_colloquium:_2010}). In recent times,
however, there has been an increasing interest in various generalization
of the Kubo response theory in various directions \citep{jin_exact_2008,uchiyama_master_2009,saeki_comparison_2010,avron_quantum_2011,wei_linear_2011,avron_adiabatic_2012,shen_hall_2014,shen_quantum_2015}.
In this paper we extend linear response theory to non-equilibrium
situations where the system's evolution is described by a time-local
master equation. %
In order to achieve this goal one has to generalize the classic  Kubo theory in two ways:
i) equilibrium thermal states are replaced by steady states of the
evolution which are, in general, non-equilibrium steady states (NESS);
ii) besides Hamiltonian perturbation, describing, for instance,
the switching on of an external field or the interaction with an external
particle, we also allow for dissipative perturbation. The latter arise
from the possibility of perturbing part of the interaction with the
environment and may become important in view of the recent developments
in the field of ``bath engineering'', according to which interaction
with a bath can be manipulated to some degree (see e.g.~\citep{kraus_preparation_2008,verstraete_quantum_2009,barreiro_open-system_2011,stannigel_constrained_2014,bardyn_topology_2013,zanardi_dissipative_2015}).
This theory can be relevant, for example, ``a little apart from thermal
equilibrium'' for weak enough system-bath coupling and/or possibly
more generally in case of engineered baths. 

At a general level, several similarities as well as differences with
the closed, unitary, response theory arise. An interesting difference
is that, for maps with a unique steady state, the open-system generalization
of the thermal susceptibility now equals the static, $\omega=0$,
susceptibility. It is well known that such quantities are in general
different in the closed case \citep{kubo_statistical-mechanical_1957}.
We also find a class of generators particularly stable against perturbations,
such that the diagonal response $\chi_{AA}(t)$ is zero for\emph{
any }Hamiltonian perturbation $A$. Such generators are Davies generators
without Hamiltonian part. Beside formulating the general theory we
also provide explicit results for several examples of dissipative
master equation. In particular we consider generalized dephasing,
Davies --thermalizing-- generators, and master equations given by
integrable, quasi-free, Majorana fermions. The latter gives us the
possibility to study linear response close a generalization of quantum
phase transitions known as dynamical phase transitions. In such transitions
it is known that the real part of the Liouvillian gap, scales to zero
at a faster rate as opposed to regular point of the phase diagram.
We show that this in turn results in a peak in the admittance $\mathrm{Im}[\hat{\chi}_{AA}(\omega)]$
at the edge of the spectrum.

\section{Dynamical response functions\label{sub:Response-functions}}



In this section we discuss, for completeness, the basic setup of response
theory for open systems. The derivations closely mirror the corresponding
ones for closed quantum systems. Similar results have been discussed
already in the literature e.g., \citep{shen_hall_2014,shen_general_2015}.

Let ${\cal H},$ denote the (finite-dimensional) Hilbert space of
the system and ${\mathrm{L}}({\cal H})$ the algebra of linear operators
over it. A time-independent Liouvillian super-operator ${\cal L}_{0}$ acting
on L$({\cal H})$ is given such that: i) 
$e^{t{\cal L}_{0}},\,(t\ge0)$  \emph{defines a semi-group of trace-preserving
positive maps with} $\|e^{t{\cal L}_{0}}\|\le1$. 

The set of \emph{steady states} of ${\cal L}_{0}$ consists of all
the quantum states $\rho$ ($\rho\in{\mathrm{L}}({\cal H}),\,\rho>0,\,{\mathrm{Tr}}\,\rho=1$)
contained in the kernel ${\mathrm{Ker}}\,{\cal L}_{0}:=\{X\,/\,{\cal L}_{0}(X)=0\}$
of ${\cal L}_{0}$. We shall denote by ${\cal P}_{0}$ (${\cal Q}_{0}:=1-{\cal P}_{0}$)
the spectral projection over Ker$\,{\cal L}_{0}$ (the complementary
subspace of Ker${\cal L}_{0}$). 
In finite dimension the condition i) implies that the non-zero
eigenvalues $\lambda_{h},\,(h>0)$ of ${\cal L}_{0}$ have non-positive
real parts. If they have strictly negative real parts then ${\cal P}_{0}=\lim_{t\to\infty}e^{t{\cal L}_{0}}$. 

We now add to ${\cal L}_{0}$ the
time-dependent perturbation $\xi_{1}(t){\cal L}_{1},$ the dynamics
is ruled by the Liouvillian ${\cal L}(t):={\cal L}_{0}+\xi_{1}(t){\cal L}_{1}$.
The evolution operator $\mathcal{E}(t)$ satisfy $d{\cal E}(t)/dt={\cal L}(t){\cal E}(t).$
We assume that the perturbation is such that 
\begin{equation}
{\cal E}(t)={\mathbf{T}}\exp\left(\int_{0}^{t}d\tau{\cal L}(\tau)\right)\label{E_t}
\end{equation}
is a family of completely positive maps. If ${\cal E}_{0}(t):=e^{t{\cal L}_{0}}$
is the one-parameter semi-group generated by ${\cal L}_{0}$ one can
write ${\cal E}(t)={\cal E}_{0}(t)+\int_{0}^{t}d\tau\xi_{1}(\tau){\cal E}_{0}(t-\tau){\cal L}_{1}{\cal E}(\tau).$
We the move to the interaction picture by defining ${\cal E}_{I}(t):={\cal E}_{0}(-t){\cal E}(t)$
which fulfills the Equation 
\begin{equation}
{\cal E}_{I}(t)=1+\int_{0}^{t}d\tau\xi_{1}(\tau){\cal L}_{1}(\tau){\cal E}_{I}(\tau),\label{E_I}
\end{equation}
where ${\cal L}_{1}(\tau):={\cal E}_{0}(-\tau){\cal L}_{1}{\cal E}_{0}(\tau)$
is the perturbation ${\cal L}_{1}$ in the interaction picture defined
by ${\cal L}_{0}$. By iteration one then finds the Born-Dyson series
for the interaction picture maps $\{{\cal E}_{I}(t)\}_{t\ge0}:$ 
\begin{widetext}
\begin{eqnarray}
{\cal E}_{I}(t)=\sum_{n=0}^{\infty}\int_{0}^{t}dt_{1}\,\xi_{1}(t_{1})\int_{0}^{t_{1}}dt_{2}\,\xi_{1}(t_{2})\cdots\int_{0}^{t_{n-1}}dt_{n}\,\xi_{1}(t_{n})\,{\cal L}_{1}(t_{1}){\cal L}_{1}(t_{2})\cdots{\cal L}_{1}(t_{n}):=\sum_{n=0}^{\infty}{\cal E}_{I}^{(n)}(t)\label{Dyson}
\end{eqnarray}

\end{widetext}

Let us now consider the time-dependent expectation value of an observable
$A$ given by $a(t):={\mathrm{Tr}}\left({\cal E}(t)(\rho)A\right)={\mathrm{Tr}}\left({\cal E}_{0}(t){\cal E}_{I}(t)(\rho)A\right),$
where $\rho$ is the system initial state. Defining $\delta a(t):=a(t)-{\mathrm{Tr}}\left({\cal E}_{0}(t)(\rho)A\right)$
and by using Eq.~ (\ref{E_I}) one obtains 
\begin{equation}
\delta a(t)=\int_{0}^{\infty}d\tau\xi_{1}(\tau)\,\chi_{NL}(t,\tau)\label{t-response}
\end{equation}
where the {\em{non-linear}} dynamical susceptibility $\chi_{NL}(t,\tau)$
is given by 
\begin{equation}
\chi_{NL}(t,\tau):=\theta(t-\tau)\,{\mathrm{Tr}}\left({\cal E}_{0}(t){\cal L}_{1}(\tau){\cal E}_{I}(t)(\rho)A\right).
\end{equation}
The latter, by resorting to Eq.~(\ref{Dyson}), can be expressed
as $\chi_{NL}(t,\tau)=\sum_{n=1}^{\infty}\chi_{NL}^{(n)}(t,\tau)$
where $\chi_{NL}^{(n)}(t,\tau):={\mathrm{Tr}}\left({\cal E}_{0}(t){\cal L}_{1}(\tau){\cal E}_{I}^{(n-1)}(t)(\rho)A\right)$
is the $n$-th order {{non-linear dynamical susceptibility}} associated
with the perturbation ${\cal L}_{1}$ and observable $A$.

The focus of this paper will be on the {\em{ linear dynamical susceptibility
}} (LDS) defined as 
\begin{equation}
\chi(t,\tau):=\chi_{NL}^{(1)}(t,\tau):=\theta(t-\tau){\mathrm{Tr}}\left({\cal E}_{0}(t){\cal L}_{1}(\tau)(\rho)A\right).\label{chi_1}
\end{equation}
Furthermore, from now on, we will assume that {\em{ the initial
state $\rho$ is a steady state of the unperturbed ${\cal L}_{0}$
i.e., ${\cal L}_{0}(\rho)=0$}}. In this case one sees that $\chi(t,\tau):=\chi_{A\mathcal{L}_{1}}(t-\tau)$
is given by 
\begin{equation}
\chi_{A\mathcal{L}_{1}}(t)=\theta(t)\,{\mathrm{Tr}}\left({\cal E}_{0}(t){\cal L}_{1}(\rho)A\right).\label{chi}
\end{equation}
In the above equation, the subscripts on $\chi$ indicates that this
is a response of the observable $A$ to the perturbation (superoperator)
$\mathcal{L}_{1}$. We will sometime omit such subscripts when the
situation is clear from the context. One can also resort to the Hilbert-Schmidt
dual maps ${\cal L}_{1}^{*}$ and ${\cal E}_{0}^{*}(t)$ and write
\begin{equation}
\chi_{A\mathcal{L}_{1}}(t)=\theta(t)\,{\mathrm{Tr}}\left(\rho\,{\cal L}_{1}^{*}(A(t))\right),\label{chi_dual}
\end{equation}
where $A(t):={\cal E}_{0}^{*}(t)(A)$ is the Heisenberg evolved $A.$
When the perturbation ${\cal L}_{1}$ is of Hamiltonian type i.e.,
${\cal L}_{1}=-i[B,\bullet],\,(B=B^{\dagger})$ we will denote the
associated LDS by $\chi_{AB}.$ In this important case the LDS becomes
\begin{equation}
\chi_{AB}(t)=i\theta(t)\,{\mathrm{Tr}}\left(\rho\,[B,A(t)]\right)=i\theta(t)\,{\mathrm{Tr}}\left([A(t),\rho]B\right),
\label{chi_unitary}
\end{equation}
Moreover if ${\cal L}_{0}$ is itself of Hamiltonian type i.e., ${\cal L}_{0}=-i[H_{0},\bullet],\,(H_{0}=H_{0}^{\dagger})$
one recovers the standard results for closed quantum systems \citep{kubo_statistical-mechanical_1957}.
In this latter case from the automorphism property ${\cal E}_{0}(t)(XY)={\cal E}_{0}(t)(X){\cal E}_{0}(t)(Y)$
of unitary evolutions and the stationarity of $\rho$ it follows that
$\chi_{AB}(t)=i\theta(t)\,{\mathrm{Tr}}\left(\rho\,[B(-t),A]\right).$
This important relation {\em{does not}} hold for a general dynamical
semi-group ${\cal E}_{0}(t).$

\subsection{Superoperator Hilbert space structures}

Given the state $\rho$ it is convenient to introduce the
(possibly degenerate) hermitean scalar product over ${\mathrm{L}}({\cal H})$ \citep{alicki_detailed_1976}
\begin{equation}
\langle A,B\rangle_{\rho}:={\mathrm{Tr}}\left(\rho A^{\dagger}B\right).\label{scalar}
\end{equation}
of $\rho={\mathbf{1}}$ one obtains the Hilbert-Schmidt scalar product
(that will be denoted simply by $\langle\bullet,\bullet\rangle$).
If ${\cal H}=i[H,\bullet],\,(H=H^{\dagger})$ and $\rho$ is any stationary
state of $H$ i.e., ${\cal H}(\rho)=0$ then the commutator map ${\cal H}$
is {\em{anti}}-hermitean with respect (\ref{scalar}). It is easy
to check that, for full-rank $\rho,$ the hermitean conjugated ${\cal M}^\sharp$ of  the linear map ${\cal M}:{\mathrm{L}}({\cal H})\mapsto{\mathrm{L}}({\cal H})$
 with respect to the scalar product (\ref{scalar}) is given by \citep{alicki_detailed_1976}
\begin{equation}
{\cal M}^\sharp(X):={\cal M}^*(X\rho) \rho^{-1}
\label{sharp}
\end{equation}
where ${\cal M}^{*}$ denotes the Hilbert-Schmidt dual map of ${\cal M},$
i.e., $\langle X,{\cal M}(Y)\rangle=\langle{\cal M}^{*}(X),Y\rangle$  \citep{alicki_detailed_1976}. In other words, with the help of the right multiplication operator  ${\cal R}_{\rho}(X)= X\rho$,  one has ${\cal M}^\sharp = {\cal R}_{\rho}^{-1} {\cal M}^* {\cal R}_{\rho}$. 
When ${\cal M}^*(X\rho)={\cal M}^*(X)\rho,\,(\forall X)$  i.e., $[{\cal M}^*,  {\cal R}_{\rho}]=0,$ one finds ${\cal M}^\sharp={\cal M}^*$ i.e., the hermitean conjugation with respect Eq.~(\ref{scalar})
coincides with the standard Hilbert-Schmidt one. This is the case, for instance, when ${\cal M}$ is a unitary group of maps and $\rho$ is one of its stationary states.
Moreover, from Eq.~(\ref{sharp}) it follows that the hermiticity  condition for $\cal M$, ${\cal  M}^\sharp={\cal M}$,  is given by
\begin{equation}
{\cal M}^{*} {\cal R}_\rho:={ {\cal R}_\rho \cal M}.\label{hermitean}
\end{equation}
Hence, if ${\cal M}$ is $\sharp$-hermitean, by applying the above equation to the state ${\mathbf{1}}$, if ${\cal M}({\mathbf{1}})={\mathbf{1}}$,
one finds ${\cal M}^{*}(\rho)=\rho$ i.e., $\rho$ is a fixed point
of the dual map ${\cal M}^{*}$. Instead if ${\cal M}({\mathbf{1}})=0$
one finds that $\rho$ is annihilated by ${\cal M}^{*}$. In the latter
case if ${\cal M}={\cal L}^{*}$ where ${\cal L}$ is a Liouvillian
the condition (\ref{hermitean}) is sometimes referred to as (generalized)
{\em{detailed balance}} and $\rho$ is a steady-state of ${\cal L}$ \citep{alicki_detailed_1976,alicki_quantum_2007}.
Notice also that, if $\rho$ is full rank, then from Eq.~(\ref{hermitean})
it follows that ${\cal M}^{*}$ is Hermitean with respect the scalar
product (\ref{scalar}) associated with $\rho^{-1}$. If the Liouvillian
${\cal L}$ is hermitean then also the dynamical maps $e^{t{\cal L}}$
are hermitean and therefore admit a spectral representation 
\begin{equation}
e^{t{\cal L}}=\sum_{\mu}e^{t\lambda_{\mu}}{\cal P}_{\mu}\label{spec_rep}
\end{equation}
where $\{\lambda_{\mu}\}_{\mu}$ are the real eigenvalues of $\cal L$
and the  superoperators ${\cal P}_\mu$ fulfill i) ${\cal P}_{\mu}{\cal P}_{\mu^{\prime}}=\delta_{\mu,\mu^{\prime}}{\cal P}_{\mu}$,
ii) $\sum_{\mu}{\cal P}_{\mu}={\mathbf{1}},$ 3) they are self hermitean
with respect (\ref{scalar}). Adding on top of such an ${\cal L}$
a commutator ${\cal H}$ such $[{\cal L},{\cal H}]=0$ and ${\cal H}(\rho)=0$
one obtain a new Liouvillian ${\cal L}^{\prime}={\cal H}+{\cal L}$
which is {\em{normal}} and therefore still admits a spectral
representation of the type (\ref{spec_rep}) but with complex $\lambda_{\mu}$'s.
This is the situation relevant to the so-called Davies generators
describing thermalization processes \citep{alicki_detailed_1976,alicki_quantum_2007}.

Using the scalar product Eq.~(\ref{scalar}) one can write Eq.~(\ref{chi_unitary}) in the following compact form 
\begin{equation}
\chi_{AB}(t)=2\theta(t)\,{\mathrm{Im}}\langle A(t),B\rangle_{\rho}.\label{chi_AB}
\end{equation}
For unitary dynamics with  Hamiltonian  $H=\sum_{n}E_{n}\Pi_{n}$, Eq.~(\ref{chi_AB}) reduces to the well known spectral formula
\begin{equation}
\chi_{AB}(t)=2\theta(t)\,{\mathrm{Im}} \sum_{n,m} e^{it(E_n - E_m)} \mathrm{Tr} (\rho \Pi_n A \Pi_m B) \label{chi_AB_unitary}
\end{equation}

\subsection{No response}
In the closed-system case it is customary e.g., in the proof of the fluctuation-dissipation theorem, to write LDS of type (\ref{chi_AB}) in terms of correlation functions $S_{AB}(t):=\langle A(t), B\rangle_\rho$ i.e., $\chi_{AB}(t)=-i\theta(t)  \left( S_{AB}(t) -S_{BA}(-t)\right).$
In the open-system case this connection cannot be established in general. In fact Eq.~(\ref{chi_AB}) can be rewritten as
\begin{equation}
\chi_{AB}(t) =-i\theta(t) \left( \langle A(t),B\rangle_{\rho} -\langle B^\sharp(t),A\rangle_{\rho}\right)
\label{diffS}
\end{equation}
where $X^\sharp(t):= ({\cal E}_{0}^{*}(t))^\sharp(X)$ fulfills the "Heisenberg-picture"  equation $dX^\sharp(t)/dt =({\cal L}_0^*)^\sharp X^\sharp(t).$
 Therefore we see that the LDS for unitary perturbations (\ref{chi_AB}) can be expressed as the difference of two correlation functions associated with 
two {\em{different}} dynamical flows.
 In the unitary case ${\cal L}_0^*=i[H_0,\bullet],$ from  $({\cal L}_0^*)^\sharp={\cal L}_0=-{\cal L}_0^*$ (see remark above),  one finds $B^\sharp(t)=B(-t)$
and the standard result is promptly recovered. 
It is interesting to notice that, when $A=B$, if the maps ${\cal E}_{0}^{*}(t)$
are hermitean with respect to the scalar product (\ref{scalar}), then Eq.~(\ref{diffS}) 
implies 
\begin{equation}
\chi_{AA}(t)=0,\quad\forall t,\quad\forall A=A^{\dagger}.\label{no_response}
\end{equation}
We will come back to this point in Sec.~\ref{sec:Davies-generators}.
This type of "diagonal'' \emph{no linear-response} for \emph{all}
observables is a uniquely open-system phenomenon. Namely, any non-trivial
unitary dynamics gives rise to a non-vanishing $\chi_{AA}$ for some
$A$. 
In fact, from Eq.~(\ref{chi_AB_unitary}), with $B=A$, 
one has 
$\chi_{AA}(t)=2 \theta(t)\sum_{n,m}\alpha_{n,m}\sin[(E_{n}-E_{m})t],
$
where $\alpha_{n,m}:={\mathrm{Tr}}\left(\rho\Pi_{n}A\Pi_{m}A\right)\in{\mathbf{R}}.$
Therefore, since $\rho$ is jointly diagonalizable with $H,$ $\chi_{AA}(t)\equiv0$
iff the observable is such that $n\neq m\Rightarrow\alpha_{n,m}=0$
namely $[A,H]=0.$ This can be true for all $A$'s iff $H$ is a scalar.

\subsection{An example: generalized dephasing}

The general Lindblad master equation has the from 
\begin{equation}
{\cal L}_{0}(\rho)=-i[H,\rho]+\sum_{\mu}\left(L_{\mu}\rho L_{\mu}^{\dagger}-\frac{1}{2}\{L_{\mu}^{\dagger}L_{\mu},\rho\}\right).\label{lindblad}
\end{equation}
Let us here consider the case in which the Lindblad operators and
the Hamiltonian are commuting with each other. In this case the set
$\{L_{\mu},\, L_{\mu}^{\dagger}\}_{\mu}\cup\{H\}$ generates a ($C^{*}$)
abelian algebra ${\cal A}$ and the kernel of $\mathcal{L}_{0}$ is
given by the commutant ${\cal A}^{\prime}$ \citep{kribs_quantum_2003}.
The Liouvillian (\ref{lindblad}) gives rise to the following family
of (dual) maps ${\cal E}_{0}^{*}(t)=e^{t{\cal L}_{0}^{*}}$ 
\begin{equation}
{\cal E}_{0}^{*}(t)(X)=\sum_{n,m}\Pi_{n}X\Pi_{m}e^{\lambda_{nm}t}\label{maps}
\end{equation}
where 1) $\{\Pi_{n}\}_{n}$ is a complete family of orthogonal projections
generating ${\cal A;}$ 2) $\lambda_{nm}=\gamma_{nm}+i\omega_{nm}$
are complex eigenvalues whose real (imaginary) parts are given by
$\gamma_{nm}=-|\gamma_{nm}|$ ($\omega_{nm}$). The condition of hermiticity
preserving and unitality implies the matrix $\Lambda=(\lambda_{nm})_{n,m}$
is hermitean with vanishing main diagonal. Plugging the expression
in Eq.~(\ref{maps}) into Eq.~(\ref{chi_AB}) one finds for $t\ge0$ 
\begin{widetext}
\begin{equation}
\chi_{AB}(t)=\bar{\chi}_{AB}+2\sum_{n\neq m}|\langle\Pi_{n}A\Pi_{m},B\rangle_{\rho}|e^{-|\gamma_{mn}|t}\sin\left(\omega_{mn}t+\theta_{nm}\right),\quad\bar{\chi}_{AB}:=-i{\mathrm{Tr}}\left([\rho,{\cal P}_{0}(A)]B\right)\label{chi_depha}
\end{equation}

\end{widetext}

where ${\cal P}_{0}(A)=\sum_{n}\Pi_{n}A\Pi_{n}\in{\cal A}^{\prime}$
is the projection of $A$ onto the kernel of ${\cal L}_{0}$ and $\theta_{nm}=\arg\,\langle\Pi_{n}A\Pi_{m},B\rangle_{\rho}$.
Notice that, since $\rho$ is a stationary state, it has the form
$\rho={\cal P}_{0}(\rho)\in{\cal A}^{\prime}$. The first, time-independent,
term in Eq.~ (\ref{chi_depha}) may be non vanishing if the commutant
algebra ${\cal A}^{\prime}$ is itself non-abelian i.e., if not all
the $\Pi_{n}$'s are rank one. The remaining terms represent a weighted
sum of response function of harmonic oscillators with resonance frequencies
(damping rates) $\omega_{nm}$ ($|\gamma_{nm}|$). Notice that if $A=B$
one has $\theta_{n,m}=0$ %
\footnote{Indeed: ${\mathrm{Tr}}\left(\rho\Pi_{m}A\Pi_{n}A\right)^{*}={\mathrm{Tr}}\left(A\Pi_{n}A\Pi_{m}\rho\right)={\mathrm{Tr}}\left(A\Pi_{n}A\rho\Pi_{m}\right)={\mathrm{Tr}}\left(\Pi_{m}A\Pi_{n}A\rho\right)={\mathrm{Tr}}\left(\rho\Pi_{m}A\Pi_{n}A\right),$
where we have used $[\Pi_{n},\rho]=0,\,(\forall n)$%
} and if moreover $\Lambda$ is real i.e., all the $\omega_{nm}$ vanish,
one has that $\chi_{AA}(t)=0\,(\forall t).$ In fact from $[\Pi_{n},\rho]=0\,(\forall n)$
it easy to check that even the first term Eq.~(\ref{chi_depha})
vanishes. This result can also be understood in light of the comment
after Eq.~(\ref{chi_AB}) by noticing that under these assumptions
the self-dual maps ${\cal E}_{0}^{*}(t)$ fulfill Eq.~(\ref{hermitean})
and are therefore hermitean.


\section{Harmonic response}

In linear response theory an important object is provided by the Fourier
transform of the LDS $\hat{\chi}_{A\mathcal{L}_{1}}(\omega):=\int dt\, e^{i\omega t}\chi_{A\mathcal{L}_{1}}(t)=\int_{0}^{\infty}dt\, e^{i\omega t}\,{\mathrm{Tr}}\left(e^{t{\cal L}_{0}}{\cal L}_{1}(\rho)A\right)$.
From this definition one readily obtains 
\begin{equation}
\hat{\chi}_{A\mathcal{L}_{1}}(\omega)=i\,{\mathrm{Tr}}\left(\frac{1}{\omega-i{\cal L}_{0}+i\varepsilon}\,{\cal L}_{1}(\rho)A\right)\label{chi_omega}
\end{equation}
where $\varepsilon=0^{+}$ is the standard regularization parameter
to make the integral above convergent (when needed, i.e.~in subspaces
where the eigenvalues of $\mathcal{L}_{0}$ are purely imaginary).
The basic response relation Eq.~(\ref{t-response}) in the $\omega$-domain
reads 
\begin{equation}
{\hat{\delta a}}(\omega):=\int dt\, e^{i\omega t}{\delta a}(t)=\hat{\xi}_{1}(\omega)\hat{\chi}_{A\mathcal{L}_{1}}(\omega).\label{omega-response}
\end{equation}
Given the assumptions on the spectrum of ${\cal L}_{0}$ one can immediately
check that $\hat{\chi}(\omega)$ is analytic in the upper $\omega$-plane
as required by causality i.e., $t<0\Rightarrow\chi(t)=0.$ Since the
${\cal E}_{0}(t)$ are hermitean preserving maps (${\cal E}_{0}(t)(X)^{\dagger}={\cal E}_{0}(t)(X^{\dagger})$)
it also easy to check that $\hat{\chi}(\omega)^{*}=\hat{\chi}(-\omega)$
and therefore the real (imaginary) part of $\hat{\chi}(\omega)$ is
a even (odd) function of $\omega.$ 

The imaginary part of of the complex susceptibility (admittance) is
known to be related to the dissipation of energy. The standard argument
still holds in this generalized setting as we now show. Let us consider
the Liouvillian with time-dependent Hamiltonian $H(t)=H_{0}+\xi(t)B$
and Liouvillian ${\cal L}=-i[H(t),\bullet]+{\cal L}_{d}.$ The time-dependent
expectation of the energy is given by $E(t)=\mathrm{Tr}\left(H(t)\rho(t)\right)$
therefore

\begin{eqnarray}
\dot{E}(t) & = & \mathrm{Tr}\left(\dot{H}(t)\rho(t)+H(t)\dot{\rho}(t)\right)\nonumber \\
 & = & \dot{\xi}(t)\mathrm{Tr}\left(B\rho(t)\right)+\mathrm{Tr}\left(H(t){\cal L}(\rho(t))\right).\label{dot-energy}
\end{eqnarray}

The last term can be written as $\dot{E}_{\mathrm{diss}}(t):={\mathrm{Tr}}\left(H(t){\cal L}_{d}(\rho(t))\right)$
and represents the energy dissipation inherently associated to the
open system dynamics ruled by ${\cal L}_{d}.$ On the other hand the, if $\rho(0)$ is a steady state
of ${\cal L}_{0}:=-i[H_{0},\bullet]+{\cal L}_{d}$, first term in Eq.~(\ref{dot-energy}) can be written
as $\dot{E}_{\mathrm{dyn}}(t):=\dot{\xi}(t)(\delta b(t)+b_{0})$ where
$b_{0}:=\mathrm{Tr}\left(B\rho(0)\right)$ and $\delta b(t):=\mathrm{Tr}\left(B\rho(t)\right)-b_{0}.$
Let us now consider a periodic perturbation $\lambda(t)\propto\cos(\Omega t).$
By averaging over a period $2\pi/\Omega$ using standard arguments
one finds

\begin{equation}
\overline{\frac{dE_{\mathrm{dyn}}(t)}{dt}}^{\frac{2\pi}{\Omega}}\propto\Omega\,{\mathrm{Im}}\,\hat{\chi}_{BB}(\Omega)
\end{equation}

We then see that the imaginary part of the Fourier-transformed LDS
accounts (only) for the energy dissipation generated by adding the
time-dependent Hamiltonian perturbation $\lambda(t)B.$ Moreover,
exactly as in the closed systems case, ${\mathrm{Im}}\,\hat{\chi}_{BB}(\Omega)$
characterizes entirely the LDS. In fact causality, even in the open-system
scenario, implies that ${\mathrm{Im}}\,\hat{\chi}_{BB}(\Omega)$ and
${\mathrm{Re}}\,\hat{\chi}_{BB}(\Omega)$ are related by the usual
Kramers-Kronig relations.

Let us now go back to a general, not necessary Hamiltonian perturbation
as in Eq.~(\ref{omega-response}). For an harmonic perturbation $\xi_{1}(t)=\cos(\Omega t),$
one finds $\delta a(t)^{\Omega}=\frac{1}{2\pi}\int d\omega\, e^{-i\omega t}\hat{\xi}_{1}(\omega)\hat{\chi}_{A\mathcal{L}_{1}}(\omega)={\mathrm{Re}}\left(e^{i\Omega t}\hat{\chi}_{A\mathcal{L}_{1}}(\Omega)\right)$
from which $|\delta a(t)^{\Omega}|\le|\hat{\chi}_{A\mathcal{L}_{1}}(\Omega)|.$
Using this inequality, Eq.~(\ref{chi_omega}) and by maximizing over
all possible normalized $A$ one gets $\sup_{\|A\|=1}|\delta a(t)^{\Omega}|\le M_{\mathrm{HR}}(\Omega)$
where 
\begin{equation}
M_{\mathrm{HR}}(\Omega)=:\|\frac{1}{\Omega-i{\cal L}_{0}+i\varepsilon}\,{\cal L}_{1}(\rho)\|_{1}\label{eq:MHR}
\end{equation}
where we also exploited the well-known inequality $|{\mathrm{Tr}}(XY)|\le\|X\|_{1}\|Y\|.$
The function $\Omega\mapsto M_{\mathrm{HR}}(\Omega)$ defined above depends on the
triple $({\cal L}_{0},{\cal L}_{1},\rho)$ but not on any observable.
The value $M_{\mathrm{HR}}(\Omega)$ sets an upper bound to the response of {\em{any}}
(normalized) observable to the perturbation ${\cal L}_{1}$ driving
harmonically at frequency $\Omega$ the system prepared in the ${\cal L}_{0}$
steady-state $\rho.$ We will refer to $M$ as the maximal harmonic
response (MHR).
\subsection{Single-qubit MHR}
To illustrate the concept of MHR we consider a single qubit subject
to the Liouvillian $${\cal L}_{0}=-i[H_{0},\bullet]+\sum_{\alpha=\pm}\gamma_{\alpha}(\sigma^{\alpha}\bullet\sigma^{-\alpha}-\frac{1}{2}\{\sigma^{-\alpha}\sigma^{\alpha},\bullet\}),$$
where $H_{0}=(\Delta/2)\sigma^{z}$ and $\gamma_{+}/\gamma_{-}=e^{-\beta\Delta}.$
The unique steady-state of ${\cal L}_{0}$ is the thermal state $\rho_{0}=\sum_{i=0,1}p_{i}|i\rangle\langle i|$
in which $p_{0}=\gamma_{-}(\gamma_{+}+\gamma_{-})^{-1}$ and $p_{1}=\gamma_{+}(\gamma_{+}+\gamma_{-})^{-1}.$
Now, if ${\cal L}_{1}=-i[B,\bullet]$ is an Hamiltonian perturbation
a straightforward computation shows that the MHR is given by 
\begin{equation}
M_{\mathrm{HR}}(\Omega)=\sum_{\alpha=\pm}\frac{|B_{01}|\tanh(\beta\Delta/2)}{|\Omega+\alpha\Delta+i\bar{\gamma}|},\label{MHR-qubit}
\end{equation}
where $\bar{\gamma}:=\frac{\gamma_{+}+\gamma_{-}}{2}$ and $B_{01}=\langle0|B|1\rangle.$
The function in Eq.~(\ref{MHR-qubit}) is a sum of two (square-root of) Lorentzians
centered at $\Omega=\pm\Delta$ with width $\bar{\gamma}$ and
maximum value $O\left(|B_{01}|\bar{\gamma}^{-1}\tanh(\beta\Delta/2)\right).$
In particular we see that for high temperature i.e., $\beta\Delta\to0$
one has $M_{\mathrm{HR}}=O(\beta\Delta)$. Of course perturbations $B$ diagonal
in the $\sigma_{z}$-basis give rise to an identically vanishing MHR.

\subsection{Relation with other susceptibilities}

Other susceptibilities, or response function, are possible. Namely
one can think to perturb the Liouvillian according to $\mathcal{L}(\lambda)=\mathcal{L}_{0}+\lambda\mathcal{L}_{1}$,
where $\mathcal{L}_{1}$ is a time-independent perturbation and $\lambda$
a (time independent) small parameter. If the system is left undisturbed
long enough, it will relax to the steady state of $\mathcal{L}(\lambda$),
$\rho(\lambda)$. If $\lambda$ is small we can ask how much the average
of an observable has changed:
\begin{align}
\langle A\rangle_{\lambda}  =\tr\left(\rho\left(\lambda\right)A\right)
  =\tr\left(\rho\left(0\right)A\right)+\lambda\chi_{A\mathcal{L}_{1}}^{T}+O(\lambda^{2}),
\nonumber
\end{align}
where $\rho=\rho(0)$ is the steady state of $\mathcal{L}_{0}$ and
we defined the out-of-equilibrium susceptibility $\chi_{A\mathcal{L}_{1}}^{T}$.
$\chi_{A\mathcal{L}_{1}}^{T}$ is the open system generalization of
the isothermal Kubo susceptibility \citep{kubo_statistical-mechanical_1957}.
Note that the state $\rho(\lambda)$ need not be thermal now. From
perturbation theory we know that (see \citep{kato_perturbation_1995})
$\rho(\lambda)=\rho-\lambda S\mathcal{L}_{1}(\rho)+O(\lambda^{2})$,
where $S$ is the reduced resolvent $S=\lim_{z\to0}Q_{0}(\mathcal{L}_{0}-z)^{-1}Q_{0}$.
Hence we obtain $\chi_{A\mathcal{L}_{1}}^{T}=-\tr\left(S\mathcal{L}_{1}(\rho_{0})A\right).$
Since $\mathcal{L}_{0}$ is a contraction semigroup we can write it
as 
\begin{equation}
\chi_{A\mathcal{L}_{1}}^{T}=\int_{0}^{\infty}dt \,\tr[Q_{0}e^{t\mathcal{L}_{0}}\mathcal{L}_{1}(\rho)A].\label{eq:chi_thermal}
\end{equation}
Comparing Eq.~(\ref{eq:chi_thermal}) with Eq.~(\ref{chi_omega})
we see that 
\begin{equation}
\hat{\chi}_{A\mathcal{L}_{1}}(0)-\chi_{A\mathcal{L}_{1}}^{T}=\int_{0}^{\infty}dt\, e^{-\varepsilon t} \tr[P_{0}\mathcal{L}_{1}(\rho_{0})B].
\end{equation}
Now, if $P_{0}$ is one-dimensional (rank-1), we have $P_{0}\mathcal{L}_{1}(\rho_{0})=\rho_{0} {\mathrm{Tr}}(\mathcal{L}_{1}(\rho_{0}))=0$
(since $\mathcal{L}_{1}(\rho_{0})$ is traceless). Hence we reach
the conclusion that, in case of non-degeneracy, the out-of-equilibrium
and the static susceptibility are equal, whereas it is well known
that this is not the case in general for the unitary case \citep{kubo_statistical-mechanical_1957}.
In case of degeneracy, instead, in general $\hat{\chi}_{A\mathcal{L}_{1}}(0)\neq\chi_{A\mathcal{L}_{1}}^{T}$,
note that, as we just said, the unitary case falls in this category.

\section{Davies generators\label{sec:Davies-generators}}

An important class of Lindbladian master equations is provided by
Davies generators \citep{davies_markovian_1974}. Such generators
arise in the limit of weak system-bath coupling and can be seen to
have Gibbs states as fixed points. A convenient generalization of
Davies generators is given by the following abstract requirements
\citep{alicki_quantum_2007}\\
\begin{enumerate}
\item The generator has the form $\mathcal{L}=\mathcal{K}+\mathcal{D}$,
where $\mathcal{K}$ is a commutator $\mathcal{K}=-i\left[H,\bullet\right]$
(and $H^{\dagger}=H$)\\
\item \emph{$\mathcal{K}^{\ast}$ }is anti-hermitean whereas $\mathcal{D}^{\ast}$
is hermitean with respect to the scalar product $\langle\bullet,\bullet\rangle_{\rho}$
(alternatively, if $\rho$ is full rank, $\mathcal{D}$ is hermitean
with respect to $\langle\bullet,\bullet\rangle_{\rho^{-1}}$)\\
\item  \emph{$\mathcal{K}$} and $\mathcal{D}$ commute. 
\end{enumerate}
The above conditions imply (together with preservation of the trace
and hermiticity) also that the state $\rho$ appearing in the scalar
product is a fixed point of the dynamics, i.e.~$\mathcal{L}(\rho)=0$.
Often one additionally imposes the so-called ergodicity, i.e.~the
requirement that $\rho$ is the unique fixed point. In other words
that, for all initial states $\rho_{0}$, $\lim_{t\to\infty}e^{t\mathcal{L}}(\rho_{0})=\rho$. 

The condition that $\mathcal{D}^{\ast}$ is hermitean with respect
to the scalar product (\ref{scalar}) has various equivalent forms.
As we have noted previously, this is equivalent to $\mathcal{D}(A\rho)=\mathcal{D}^{\ast}(A)\rho$.
Another equivalent formulation is that the map defined by $\mathcal{F}=({\cal R}_{\rho^{1/2}})^{-1}\mathcal{D} {\cal R}_{\rho^{1/2}}$,
is hermitean according to the standard Hilbert-Schmidt scalar product. Explicitly,
$\tr(A^{\dagger}\mathcal{F}(B))=\tr\left(\left[\mathcal{F}(A)\right]^{\dagger}B\right)$
with $\mathcal{F}(x)=\mathcal{D}(x\rho^{1/2})\rho^{-1/2}$. This observation
implies at once that such generators $\mathcal{D}$ have a purely
real spectrum. 

As noted in Sec.~\ref{sub:Response-functions}, this is precisely
the condition leading to $\chi_{AA}(t)=0$ for all $A$. In other
words, purely dissipative Davies generators, i.e.~for which $\mathcal{K}=0$,
are very stable to perturbations in that the linear, diagonal, response
function $\chi_{AA}(t)$ vanish identically for all $A$. 

\subsection{Single qubit}
In the following we will consider in detail the qubit case. The most
general Davies map for the single qubit has been characterized in
full detail in \citep{roga_davies_2010}. Setting the quantization
axis along the basis of the Hamiltonian, the most general Davies map,
in the Schr\"{o}dinger representation $\phi_{t}=e^{t\mathcal{L}}$,
has the following matrix form
\begin{equation}
\phi_{t}=\left(\begin{array}{cccc}
1-a & 0 & 0 & a\frac{p}{1-p}\\
0 & ce^{-it\Delta} & 0 & 0\\
0 & 0 & ce^{it\Delta} & 0\\
a & 0 & 0 & 1-a\frac{p}{1-p}
\end{array}\right).\label{eq:davies}
\end{equation}
The parameters are given by $a=(1-p)(1-e^{-bt})$ and $c=e^{-\Gamma t}$,
whereas the unique fixed point is given by the density matrix
\begin{equation}
\rho=\left(\begin{array}{cc}
p & 0\\
0 & 1-p
\end{array}\right),
\end{equation}
and $p=e^{\Delta/(2T)}/[2\cosh(\Delta/2T)]$ for a temperature $T$,
such that $\rho$ is in Gibbs form. The conditions that $\phi_{t}$
is a valid, completely positive, map are $0\le p\le1$ and $0\le a\le1$.
Moreover the rates must satisfy the condition $\Gamma\ge b/2\ge0$
\citep{roga_davies_2010}. With respect to Ref.~\citep{roga_davies_2010}
we included the effect of a Hamiltonian term with $H=\Delta\sigma^{z}/2$.
Without such a Hamiltonian term, for what said previously, one would
always have $\chi_{AA}(t)=0$. Using explicitly Eq.~(\ref{eq:davies})
(or rather its adjoint), it is not difficult to compute the general,
linear, response to a Hamiltonian perturbation. The result is 
\begin{equation}
\chi_{AB}(t) = 
  2\theta(t)e^{-t\Gamma}\sin(t\Delta+\varphi)\tanh(\frac{\Delta}{2T})\left|A_{01}B_{10}\right|,
\end{equation}
where $\varphi=\arg(A_{01}B_{10})$. One can explicitly check that
$\chi_{AA}(t)=0$ for the purely dissipative case as then one has
$\varphi=\Delta=0$. Its Fourier transform has a like-wise familiar
form
\begin{equation}
\hat{\chi}_{AB}(\omega)=\tanh(\frac{\Delta}{2T})\left|A_{01}B_{10}\right|\Big[\frac{e^{i\varphi}}{\omega+\Delta+i\Gamma}-\frac{e^{-i\varphi}}{\omega-\Delta+i\Gamma}\Big]
\end{equation}

We now consider more general perturbations which cannot be written
as a commutator. We consider perturbations which are themselves Davies
generator. Using Eq.~(\ref{chi_dual}) we express the LDS as 
\begin{equation}
\chi_{A\mathcal{L}_{1}}(t)=\theta(t)\left.\frac{d}{ds}\right|_{s=0}\tr\left[\rho e^{s\mathcal{L}_{1}^{\ast}}(\mathcal{E}_{t}^{\ast}(A))\right].
\end{equation}
We will consider $\mathcal{L}_{1}$ to be the most general Davies
generator, but of course in another direction with respect to $\mathcal{E}_{t}$.
We then consider the rotated version of a Davies map $\phi^{U}$,
$\phi^{U}=U^{\dagger}\otimes U^{T}\,\phi\, U\otimes\overline{U}$
where $U$ is a $SU(2)$ matrix that empowers a rotation in the Hilbert
space (we exploited here the isomorphism between superoperator space
and $\mathcal{H}\otimes\mathcal{H}$). 

Now we write the most general $SU(2)$ matrix as $U=e^{i\alpha\hat{n}\cdot\boldsymbol{\sigma}/2}$
$(\hat{n}=(n_{x},n_{y},n_{z})$). Explicitly 
\begin{equation}
U=\left(\begin{array}{cc}
\cos\left(\frac{\mbox{\ensuremath{\alpha}}}{2}\right)+in_{z}\sin\left(\frac{\mbox{\ensuremath{\alpha}}}{2}\right) & (in_{x}+n_{y})\sin\left(\frac{\mbox{\ensuremath{\alpha}}}{2}\right)\\
(in_{x}-n_{y})\sin\left(\frac{\mbox{\ensuremath{\alpha}}}{2}\right) & \cos\left(\frac{\mbox{\ensuremath{\alpha}}}{2}\right)-in_{z}\sin\left(\frac{\mbox{\ensuremath{\alpha}}}{2}\right)
\end{array}\right)
\end{equation}

To get a grasp we take $\hat{n}=(1,0,0)$, i.e.~a rotation of $\alpha$
around the $x$ axis. We also remove the Hamiltonian part from $\mathcal{L}_{1}$.
The result is
\begin{widetext}
\begin{eqnarray}
\chi_{A\mathcal{L}_{1}}(t) & = & \frac{\theta(t)}{4}\Bigg\{ e^{-bt}\left(A_{11}-A_{00}\right)\left\{ \left[(b_{1}+\Gamma_{1}+(b_{1}-\Gamma_{1})\cos(2\alpha)\right]\tanh\left(\frac{\Delta}{2T}\right)-2b_{1}\cos\left(\alpha\right)\tanh\left(\frac{\Delta_{1}}{2T_{1}}\right)\right\} \nonumber\\
 &  & -2e^{-t\Gamma}\left|A_{01}\right|\sin(t\Delta+\varphi_{01})\left[(b_{1}-\Gamma_{1})\sin(2\alpha)\tanh\left(\frac{\Delta}{2T}\right)-2b_{1}\sin\left(\alpha\right)\tanh\left(\frac{\Delta_{1}}{2T_{1}}\right)\right]\Bigg\},
\end{eqnarray}

\end{widetext}

with $\varphi_{01}=\arg(A_{01})$. A rotation around $y$ results
in a very similar expression.

\section{Open quasi-free systems}

For integrable Hamiltonians, quadratic in creation and annihilation
operators, the Lindblad master equation (and in fact even more general
version thereof) is solvable provided that the Lindblad operators
$L_{\mu}$ appearing in Eq.~(\ref{lindblad}) are \emph{linear} in
creation/annihilation operators. The solvability of such master equations
was first proved in \citep{prosen_third_2008} and later investigated
in series of work (see e.g.~\citep{prosen_quantum_2008,prosen_spectral_2010,prosen_quantization_2010,zunkovic_explicit_2010,eisert_noise-driven_2010,znidaric_solvable_2011,horstmann_noise-driven_2013,banchi_quantum_2014,znidaric_exact_2014,znidaric_relaxation_2015}
for a non-comprehensive list of references). In this section we are
going to present a detailed analysis of linear response functions
for such open, quasi-free systems. For concreteness we focus on Fermi
systems. It is convenient to encode such problems in terms of Majorana
operators, the Fermionic analogue of positions and momenta. For Fermi
operators $f_{i},f_{j}^{\dagger}$ satisfying $\left\{ f_{i},f_{j}^{\dagger}\right\} =\delta_{i,j}$
we define the following Majorana operators $m_{1,j}=f_{j}+f_{j}^{\dagger}$,
$m_{2,j}=i(f_{j}-f_{j}^{\dagger})$ such that $\{m_{\lambda,i},m_{\gamma,j}\}=2\delta_{\lambda,\gamma}\delta_{i,j}$.
Often we will use a single multi-index $i$ in place of $i,\lambda$.
Quadratic, hermitean observables, such as the Hamiltonian, can be
written as 
\begin{equation}
H=\Gamma(\mathsf{H})=\frac{i}{4}\sum_{i,j}\mathsf{H}_{i,j}m_{i}m_{j}=\frac{i}{4}\boldsymbol{m}\mathsf{H}\boldsymbol{m},\label{eq:gamma_H}
\end{equation}
where we also employed a matrix-vector notation. Here $\mathsf{H}\in\mathbb{R}^{2L\times2L}$
is a $2L\times2L$ real antisymmetric matrix, $\mathsf{H}^{T}=-\mathsf{H}$
and $L$ is the number of fermionic modes. Remind that Gaussian states
are those state which satisfy a Wick theorem. For open, quasi-free,
Lindblad generators, $H$ has the form of Eq.~(\ref{eq:gamma_H})
and the Lindblad operators can be written as $L_{\mu}=\sum_{i}\mathsf{L}_{i}^{\mu}m_{i}$.
The corresponding evolution operator, $\mathcal{E}_{t}=e^{t\mathcal{L}}$,
maps Gaussian states to Gaussian states. Gaussian states $\rho$ are
uniquely characterized by their covariance matrix, given by
\begin{equation}
C_{i,j}=\frac{i}{2}\tr\left(\rho\left[m_{i},m_{j}\right]\right)=-\mathrm{Im}\langle m_{i}m_{j}\rangle,
\end{equation}
which, with this convention, is also real and anti-symmetric. It should
then be possible to map the Lindblad equation (\ref{lindblad}) into
an equation for the covariance matrix $C$. This is indeed the case
\citep{eisert_noise-driven_2010,horstmann_noise-driven_2013}, and
one can show that $C$ satisfies the following affine differential
equation
\begin{equation}
\dot{C}=XC+CX^{T}-Y.\label{eq:affine}
\end{equation}
The matrices $X$ and $Y$ are given by
\begin{align}
X & =H-S\\
Y & =2i\sum_{\mu}|\mathsf{L}^{\mu}\rangle\langle\mathsf{L}^{\mu}|-|\mathsf{L}^{\mu\ast}\rangle\langle\mathsf{L}^{\mu\ast}|\\
S & =\sum_{\mu}|\mathsf{L}^{\mu}\rangle\langle\mathsf{L}^{\mu}|+|\mathsf{L}^{\mu\ast}\rangle\langle\mathsf{L}^{\mu\ast}|
\end{align}
 where we indicated with $|\mathsf{L}^{\mu}\rangle$ the $2L\times1$
vector of components $L_{i}^{\mu}$ and $\langle\mathsf{L}^{\mu}|$
its complex conjugate transpose. Hence both $S$ and $Y$ are real
matrices. Moreover $H$ is the antisymmetric part of $X$ whereas
$-S$ is its symmetric part. Since $S$ is positive semi-definite
($S\ge0$), it follows that the eigenvalues of $X$ have non-positive
real part \citep{prosen_spectral_2010,zunkovic_explicit_2010,horstmann_noise-driven_2013}. 

We now turn to the computation of linear response for such quasi-free
open system. We assume that both the unperturbed generator and the
perturbation can be written as a quasi-free open system generator
in the way we just specified. We use a similar trick as the one in
section \ref{sec:Davies-generators} but now we start from Eq.~(\ref{chi}).
In other words we write
\begin{equation}
\chi_{A\mathcal{L}_{1}}(t)=\theta(t)\left.\frac{d}{ds}\right|_{s=0}\mathrm{Tr}\left([{\cal E}_{0}(t)e^{s{\cal L}_{1}}](\rho)A\right)\label{eq:chi_quasifree}
\end{equation}
Now, for quasi-free evolutions $\mathcal{E}_{t}$, and Gaussian state
$\rho_{C_{0}}$ with covariance matrix $C_{0}$, we have $\mathcal{E}_{t}(\rho_{C_{0}})=\rho_{C_{t}}$
where $C_{t}$ is the solution of the corresponding differential equation
(\ref{eq:affine}) with initial value $C_{0}$. Let us indicate with
$g_{t}^{\alpha}$ the flow corresponding to the differential equation
(\ref{eq:affine}) with $\alpha=0$ corresponding to the unperturbed
generator and $\alpha=1$ for the perturbed one. Hence 
\begin{equation}
\mathcal{E}_{0}(t)[e^{s\mathcal{L}_{1}}(\rho_{C_{0}})]=\mathcal{E}_{t}\left(\rho_{g_{s}^{1}(C_{0})}\right)=\rho_{g_{t}^{0}(g_{s}^{1}(C_{0}))}.
\end{equation}

With the help of the super-operator $\hat{X}_{\alpha}(C):=X_{\alpha}C+CX_{\alpha}^{T}$,
the equation $\dot{C}=X_{\alpha}C+CX_{\alpha}^{T}-Y_{\alpha}$ becomes
$\dot{C}=\hat{X}_{\alpha}C-Y_{\alpha}$ whose solution is
\begin{equation}
C(t)=e^{t\hat{X}_{\alpha}}[C(0)]-\int_{0}^{t}e^{(t-\tau)\hat{X}_{\alpha}}[Y_{\alpha}]d\tau.
\end{equation}
 Using equation (\ref{eq:chi_quasifree}) we can now compute the LDS
for any $n$-point operator $A$ using Wick theorem. For simplicity
we stick to the case where $A$ is quadratic, i.e.~$A=\Gamma(\mathsf{A})$.
Then 
\begin{equation}
\tr\left[[\mathcal{E}_{0}(t)\circ e^{s\mathcal{L}_{1}}](\rho)\Gamma(\mathsf{A})\right]=\frac{1}{4}\tr\left[\mathsf{A}^{T}g_{t}^{0}(g_{s}^{1}(C_{0}))\right].
\end{equation}
We can now take the derivative with respect to $s$, noting that,
clearly, 
\begin{equation}
\left.\frac{d}{ds}g_{s}^{1}(C_{0})\right|_{s=0}=X_{1}C_{0}+C_{0}X_{1}^{T}-Y_{1}.
\end{equation}
Defining $C_{1}=X_{1}C_{0}+C_{0}X_{1}^{T}-Y_{1}$, we finally obtain
\begin{eqnarray}
\chi_{A\mathcal{L}_{1}}(t) & = & \frac{\theta(t)}{4}\tr\left[\mathsf{A}^{T}e^{t\hat{X}_{0}}(C_{1})\right].\label{eq:chi_QF_final}
\end{eqnarray}
The above equation is the one-particle analogue of Eq.~(\ref{chi}).
Note however that the expression Eq.~(\ref{eq:chi_QF_final}) is
independent of $Y_{0}$. Fourier transforming we obtain the analog
of Eq.~(\ref{chi_omega}) in the quasi-free setting: 
\begin{equation}
\hat{\chi}_{A\mathcal{L}_{1}}(\omega)=i\frac{1}{4}\tr[\mathsf{A}^{T}\frac{1}{\omega-i\hat{X}_{0}+i\epsilon}(C_{1})].\label{eq:chi_om_QS}
\end{equation}
If the matrix $X$ can be diagonalized, i.e.~there exist a non-singular
matrix $V$ such that $V^{-1}XV=\mathrm{diag}\left\{ \xi_{1},\ldots,\lambda_{2L}\right\} $,
a convenient expression for the LDS is given by
\begin{equation}
\hat{\chi}_{A\mathcal{L}_{1}}(\omega)=i\frac{1}{4}\sum_{k,q}\frac{1}{\omega-i(\lambda_{k}+\lambda_{q})+i\epsilon}[\hat{\mathsf{A}^{T}}]_{k,q}[\check{C}_{1}]_{q,k}\,,\label{eq:chi_V}
\end{equation}
with $\hat{\mathsf{A}^{T}}=V^{T}\mathsf{A}^{T}V$ and $\check{C}_{1}=V^{-1}C_{1}(V^{T})^{-1}$.
Using the same argument to arrive at Eq.~(\ref{eq:MHR}) applied
on Eq.~(\ref{eq:chi_om_QS}) we can obtain a single-particle analog
of the MHR. Namely
\begin{equation}
\mathsf{M}_{\mathrm{HR}}(\omega):=\frac{1}{4}\left\Vert \frac{1}{\omega-i\hat{X}_{0}+i\epsilon}(C_{1})\right\Vert _{1}.\label{eq:single_MHR}
\end{equation}

\begin{figure}
\begin{centering}
\includegraphics[width=7cm,height=4cm]{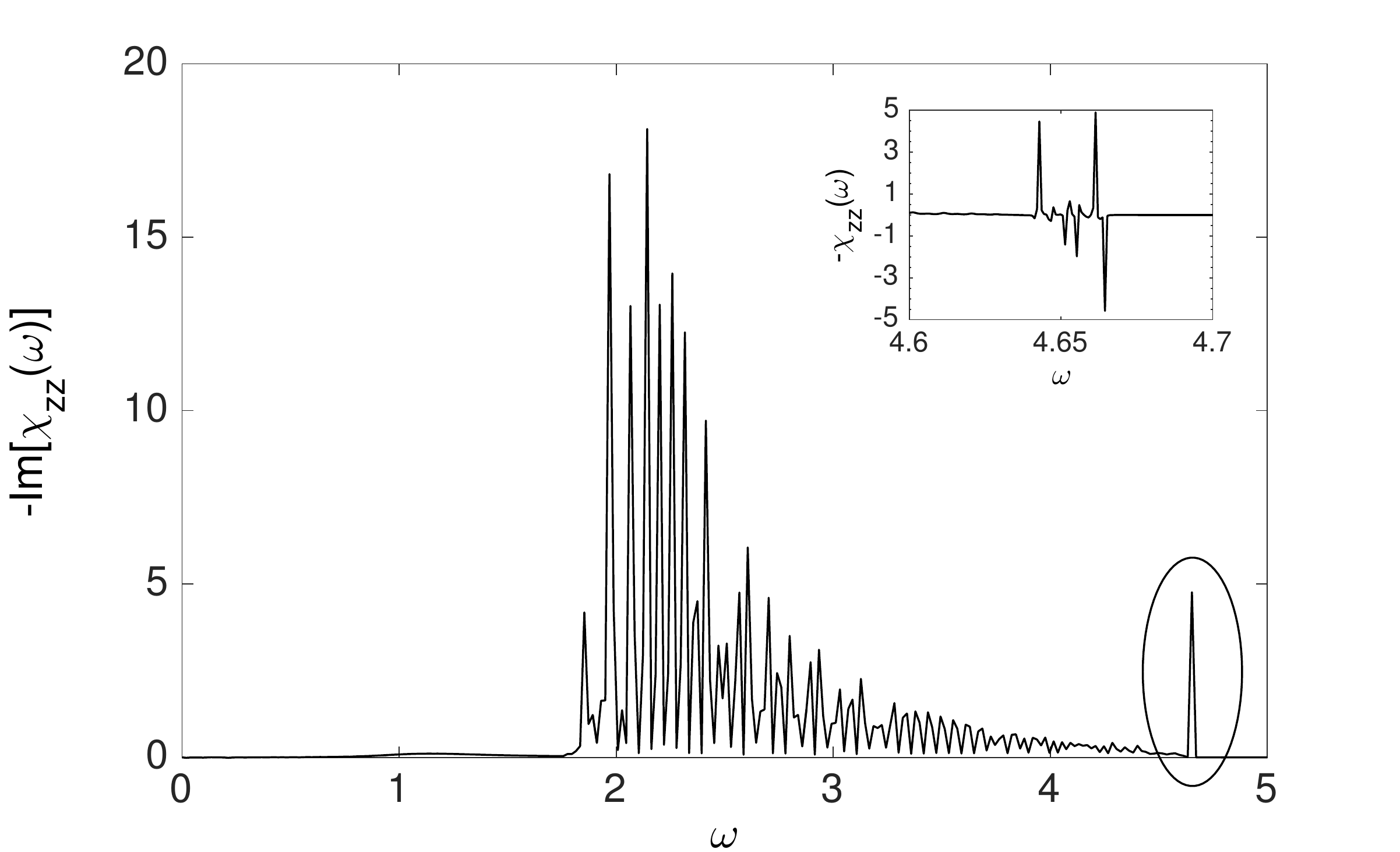}
\par\end{centering}

\begin{centering}
\includegraphics[width=7cm,height=5cm]{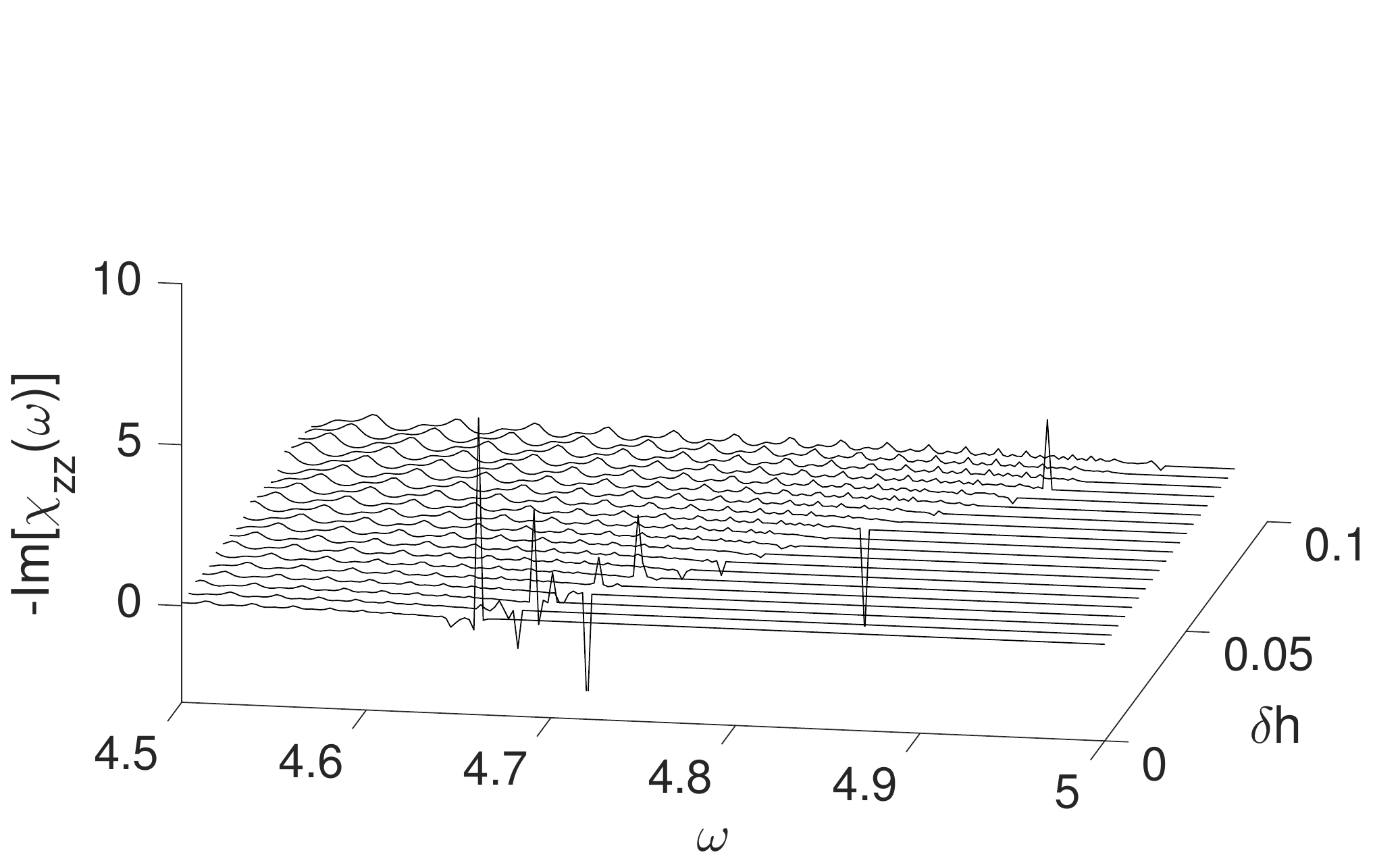}
\par\end{centering}

\protect\caption{Top panel: $\mathrm{Im}(\chi(\omega))$ for a chain of $N=200$ sites
at a non-equilibrium critical point. The ellipsis emphasizes the peak
related to the closing of the Liouvillian gap $\Delta$. In the inset
a zoom of the region around the peak shows the non-monotonic behavior
of the peak with $\omega$. Bottom panel, a region of frequencies
around the peak location is shown for different fields $\delta h$
close to the critical point $h=h_{c}+\delta h$ (with $h_{c}=1-\gamma^{2}$) for $N=200$ sites.
Peaks are presents also sufficiently close to the critical point however
there is a non-monotonic behavior as described in the main text. Other parameters are $\gamma = 0.8$, $\Gamma_1^L = 0.3, \, \Gamma_2^L = 0.1$, $\Gamma_1^R = 0.3, \, \Gamma_2^R = 0.2$. \label{fig:LDS-peak}}
\end{figure}

\subsection{Dissipative $XY$ spin-chain}

We will now study a specific example of quasi-free open system. Namely
we consider the model introduced in \citep{prosen_third_2008,prosen_quantum_2008}
of an $XY$ spin-chain with thermal magnetic baths acting at the ends
of the chain. The model is given by Eq.~(\ref{lindblad}) with Hamiltonian
\[
H=\sum_{n=1}^{N-1}\left(\frac{1+\gamma}{2}\sigma_{n}^{x}\sigma_{n+1}^{x}+\frac{1-\gamma}{2}\sigma_{n}^{y}\sigma_{n+1}^{y}\right)+\sum_{n=1}^{N}h\sigma_{n}^{z}\,,
\]
and the following four Lindblad operators
\begin{align*}
L_{1}=\sqrt{\Gamma_{1}^{L}}\sigma_{1}^{-}, & \,\, L_{3}=\sqrt{\Gamma_{1}^{R}}\sigma_{N}^{-},\\
L_{2}=\sqrt{\Gamma_{2}^{L}}\sigma_{1}^{+}, & \,\, L_{4}=\sqrt{\Gamma_{2}^{R}}\sigma_{N}^{+},
\end{align*}
where $\sigma_{n}^{\pm}=(\sigma_{n}^{x}\pm i\sigma_{n}^{y})/2$ and
$\Gamma_{1,2}^{L,R}$ are positive constants related to the baths
temperature at the ends. Namely $\Gamma_{2}^{\ell}/\Gamma_{1}^{\ell}=\exp(-2h/T_{\ell})$
for $\ell=L,R$. In the thermodynamic limit the model has a critical
``line'' $h_{c}^{2}=(1-\gamma^{2})^{2}$ where the correlations
$\langle\sigma_{n}^{z}\sigma_{n+r}^{z}\rangle$ decay algebraically
as $r^{-4}$. Outside criticality the decay is exponential with a
correlation length $\xi=1/4\mathrm{arccosh}(h/h_{c})]$, hence $\xi$
diverges with a mean-field exponent, $\xi\sim\left|h-h_{c}\right|^{-1/2}$,
as the critical point is approached. 

The matrix $X$, as can be seen
from Eq.~(\ref{eq:affine}), plays an analogous role as the Hamiltonian
in this open system setting. In our numerical simulations we verified
that $X$ could always be diagonalized. In this case the Lindblad
generator can be cast in the following normal form $\mathcal{L}_{0}=\sum_{k}\lambda_{k}d_{k}^{\times}d_{k}$
where the operators $d_{k}^{\times}\neq d_{k}^{\dagger}$ but otherwise
satisfy canonical anticommutation relations such that $d_{k}^{\times}d_{k}$
are non-hermitean number operators. 
As proven in \citep{prosen_third_2008}
the non-equilibrium steady state is unique iff $\lambda_{k}\neq0$
for all $k$. In this case the convergence to the NESS is exponential
with a rate given by the ``gap'' $\Delta=2\min_{k}\{-\mathrm{Re}(\lambda_{k})\}$.
Such non-equilibrium phase transition are also characterized by a
different scaling to zero with $N$ of the dissipative gap $\Delta$
\citep{prosen_quantum_2008}. Namely, as $N\to\infty$, one has $\Delta(N)\sim N^{-5}$
at the critical points whereas $\Delta(N)\sim N^{-3}$ elsewhere.

\begin{figure}
\begin{centering}
\includegraphics[width=7cm,height=5cm]{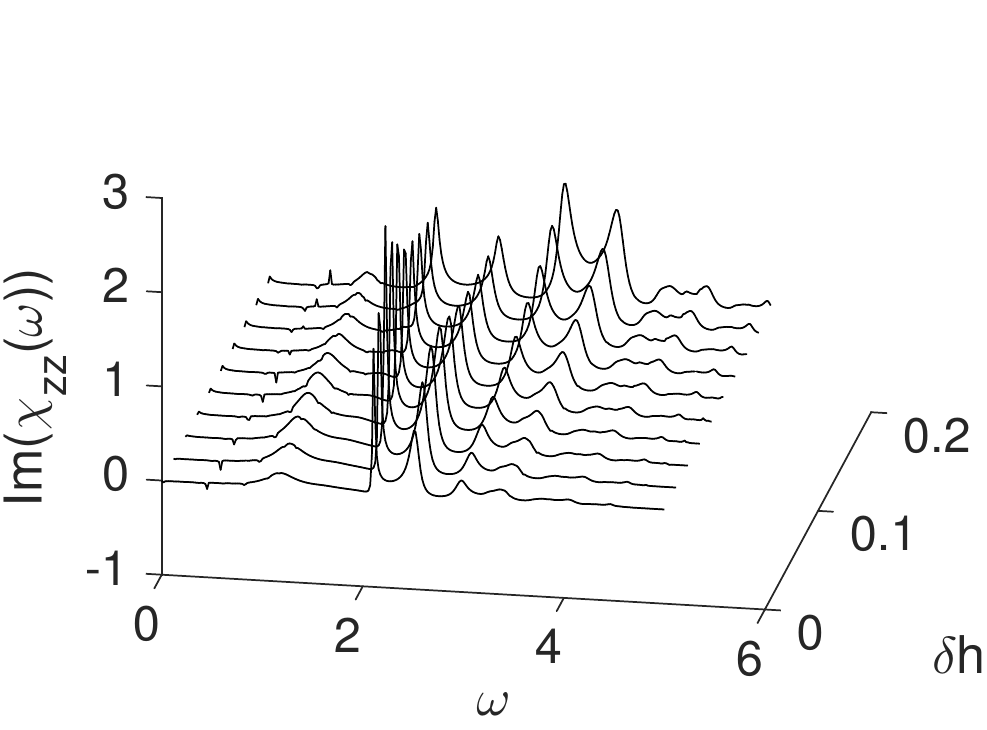}
\par\end{centering}

\begin{centering}
\includegraphics[width=7cm,height=5cm]{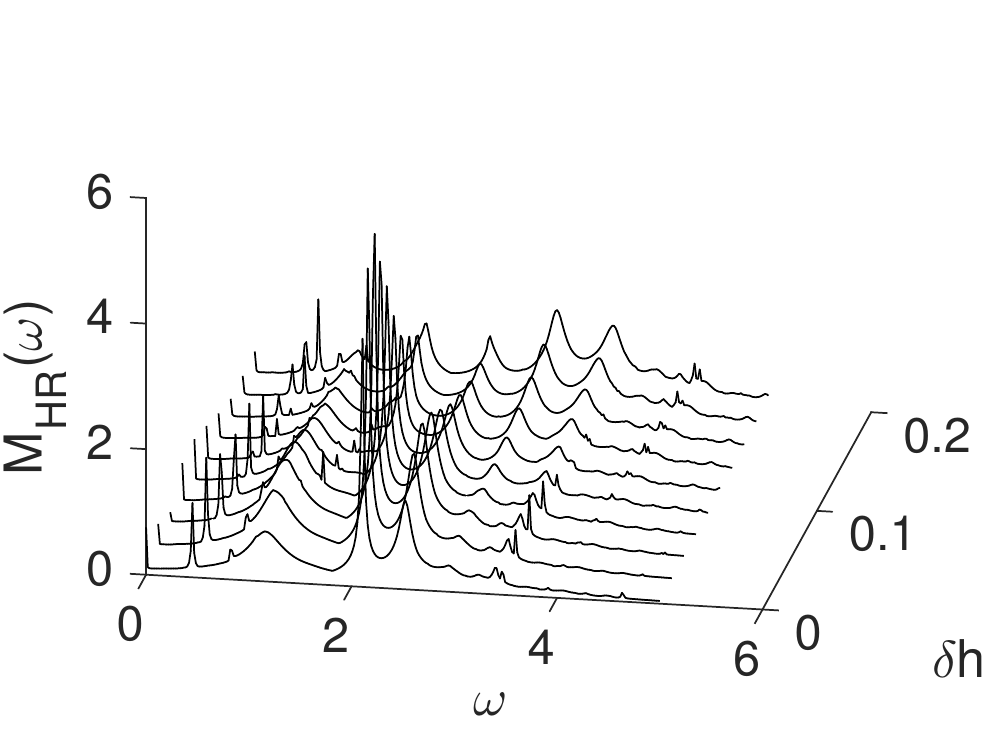}
\par\end{centering}

\protect\caption{Comparison between LDS and the single particle MHR Eq.~(\ref{eq:single_MHR}).
Top panel: imaginary LDS in frequency domain computed using Eq.~(\ref{eq:chi_V}).
Bottom panel: single particle MHR $\mathsf{M}_{\mathrm{HR}}(\omega)$ as given by
Eq.~(\ref{eq:single_MHR}). The system's size is $N=14$ and the
external field is $h=h_{c}+\delta h$, where $h_{c}=1-\gamma^{2}$
is the critical field. Other parameters as in Fig.~\ref{fig:LDS-peak}.
\label{fig:comparison}}
\end{figure}

\begin{figure}
\begin{centering}
\includegraphics[width=7cm,height=5cm]{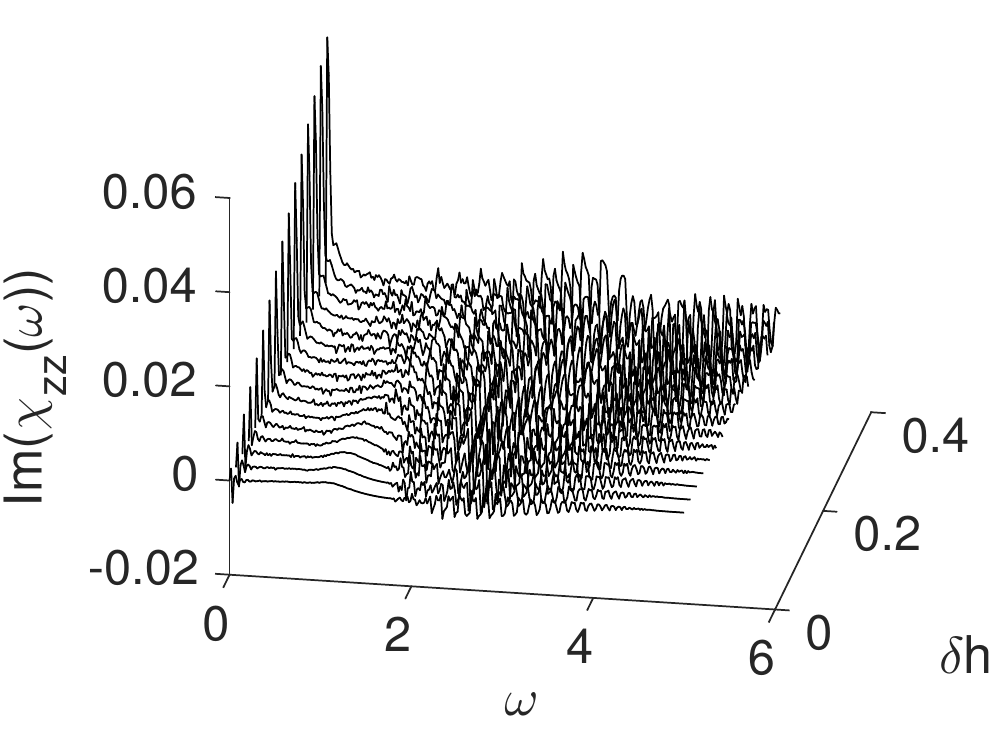}
\par\end{centering}

\protect\caption{LDS for a purely dissipative perturbation. Size is $N=100$. Other parameters as in Fig.~\ref{fig:LDS-peak}. \label{fig:LDS_dissi} }
\end{figure}

At the critical point we numerically observe a multitude of levels
whose real part is going to zero. This seems to be analogous to standard,
Hamiltonian, second order phase transitions where an extensive number
of energy gaps go to zero in the thermodynamic limit. In our simulations
we observe this feature at the edge of the spectrum i.e.~for $\omega\approx\pm2\max\mathrm{Im}(\lambda_{k})$.
Assume then, that for a certain number of $k$'s one has, approximately
$\lambda_{k}\approx\sigma-i\rho$ with $|\sigma|\ll1$. The denominator
in Eq.~(\ref{eq:chi_V}) gives rise to a contribution of the form
$[2\sigma-i(\omega-2\rho)]/[(\omega-2\rho)^{2}+4\sigma^{2}]$, in
other words we expect a strong, Lorentzian, peak at $\omega\approx2\rho$.
This argument finds indeed numerical confirmation as can be seen from
figure \ref{fig:LDS-peak}. Such peaks are present in a quasi-critical
region, sufficiently close to the (out-of-equilibrium) critical point.
The size and scaling properties of such peaks are, however, difficult
to predict. For example the peak hight is not necessarily increasing
with systems size. The reason is that the numerator $[\hat{\mathsf{A}^{T}}]_{k,q}[\check{C}_{1}]_{q,k}$
in Eq.~(\ref{eq:chi_V}) not only does not have a definite sign but
is in fact complex. The overall contribution to $\mathrm{Im}(\chi(\omega))$
is a linear combination of peaked Lorentzian with coefficients of
possibly different sign. This effect can be appreciated in Fig.~\ref{fig:LDS-peak}
bottom panel where peaks are shown for a region of field close to
criticality. As a function of the external field, peaks change sign
and may even disappear completely as a consequence of destructive
interference. 
In Fig.~\ref{fig:comparison} we also plot the single particle MHR,
Eq.~(\ref{eq:single_MHR}) and compare it with the LDS. The MHR reveals similar features as $\mathrm{Im}[\chi_{zz}(\omega)]$ albeit possibly more pronounced. 
Finally we consider perturbation of purely dissipative character,
i.e.~we set $\mathcal{L}_{1}=\mathcal{L}_{0}+i\left[H,\bullet\right]$.
The results are shown in Fig.~\ref{fig:LDS_dissi}


\section{Conclusions}

In this paper we discussed the extension of   Kubo linear response theory to open quantum systems
whose dynamics is described by a master equation generating
a semi-group of contractive dynamical maps. 

The theory parallels the standard  closed
case but some important differences arise. For example, for generators with
a unique steady state, the generalization of the thermal susceptibility
becomes now equal to the $\omega=0$ complex admittance. This is known
not to be the case in the unitary setting \citep{kubo_statistical-mechanical_1957}.
Moreover for a class of hermitean dynamical maps we have  shown that the diagonal response functions
are identically vanishing.
We derived exact expressions for the linear dynamical response functions for generalized dephasing, Davies generators,
and integrable, quasi-free master equation. We introduced the observable-free notion of maximal harmonic response
and computed it explicitly for a single qubit.
 
In  the quasi-free case we concentrated the analysis close to the dynamical phase transition points
which are known to take place in these systems. It is found that a
signature of such dynamical phase transitions shows up as a peak in
the imaginary part of the admittance at the edge of the spectrum.

Applications of our  dynamical response theory to a variety of physically relevant systems as well as its  extension  to wider class of open quantum system dynamics e.g., non-Markovian,  clearly  deserve future investigations.
\begin{acknowledgments}
This work was partially supported by the ARO MURI Grant No.  W911NF-11-1-0268
\end{acknowledgments}

\bibliographystyle{apsrev4-1}
\bibliography{linear_response}

\end{document}